\documentstyle[12pt,oldlfont]{article}
\normalbaselines
\begin{document}
\bibliographystyle{unsrt}

\vspace{4cm}

\vbox{\vspace{38mm}}
{\hfill Preprint INFN-NA-IV-94/33~~~~~~~~~~~~~~~~~~~~~~~~~~~DSF-T-94/33}

\title{            STATISTICS OF PARAMETRICALLY EXCITED
                         PHOTON-ADDED COHERENT STATE           }

\author{   V.V.Dodonov, Ya.A.Korennoy, \\
         V.I.Man'ko$^*$, and Y.A.Moukhin           \\
         Lebedev Physics Institute,                          \\
       Leninsky prospect 53, 117924 Moscow, Russia                   \\
	$^*$ $Lebedev Physics Institute$ $and$ \\
       $Dipartimento$ $di$ $Scienze$ $Fisiche,$	\\
	$Universita$ $di$ $Napoli$ $``Federico II'',$  		\\
       $Mostra$ $d'Oltremare,$ $Pad.$ $19-80125$ $Napoli,$ $Italy$  }

\maketitle
\newcommand{\be}{\begin{equation}}
\newcommand{\ee}{\end{equation}}
\newcommand{\ds}{\displaystyle}
\newcommand{\bdm}{\begin{displaymath}}
\newcommand{\edm}{\end{displaymath}}

\newcommand{\goo}{\,\raisebox{-.5ex}{$\stackrel{>}{\scriptstyle\sim}$}\,}
\newcommand{\loo}{\,\raisebox{-.5ex}{$\stackrel{<}{\scriptstyle\sim}$}\,}
\section*{ Abstract }
Photon distribution function,means and dispersions are found explicitly
for the nonclassical state of light which is created from the photon--added
coherent state $\vert \alpha,m \rangle$
due to a time--dependence of the frequency
of the electromagnetic field oscillator. Generating function for
factorial momenta is obtained. The Wigner function and Q--function
are constructed explicitly for the excited photon--added coherent state
of light. Influence of added photons on known oscillations of photon
distribution function for squeezed light is demonstrated.

\hspace{1.5em}
\newpage
\section{ Introduction}
%
There are several types of nonclassical states of one--mode light,
statistics of which has been studied last decade. They include squeezed
light \cite{P2}, \cite{Nieto}, \cite{P5}, correlated light \cite{P14},
even and odd coherent states \cite{Physica},
\cite{Dodonov.Man'ko.Nikonov} or Schr\"odinger
cat states \cite{Yurke.Stoler}. All these nonclassical states demonstrate
either sub--poissonian or super--poissonian statistics of photons.
The statistics of the coherent photons \cite{Glauber}, \cite{Klauder},
\cite{Sudarshan}
is described by the Poisson distribution function. The coherent states of
light are considered as the classical states.

The squeezed light has the property of the reducing the dispersion in
one of the photon quadrature components. The photon distribution function
of squeezed light has wavy behaviour for large squeezing \cite{Sch.W},
\cite{Vourdas.Weiner}.
The quadrature components of the squeezed light satisfy the condition of
the minimization of the Heisenberg uncertaintly relation \cite{PL1}.

The correlated states of light \cite{P14} satisfy the condition of the
minimization of Schr\"odinger uncertaintly relation \cite{P15}, \cite{P16}.
The photon distribution function of the correlated light has the
oscillatory behaviour \cite{P17}, \cite{P18} depending on the correlation
coefficient of the quadrature components which is extra physical
parameter characterizing the state of the electromagnetic field
oscillator. The property of oscillations of the photon distribution
function is presented also for two--mode light \cite{P19}, \cite{P20},
\cite{P21}.

In Ref.\cite{Ag.T.} the state $\vert \alpha,m \rangle$, which is
obtained by the multiple action of the creation operator on the usual
coherent state has been studied. The photon--added coherent state
$\vert  \alpha ,m\rangle$ depends on extra discrete parameter, number
of excited photons, which influences the statistics of photons.
The statistics of this light including distribution function,
means and dispersions of quadrature components, Wigner function and
$Q$--function were found explicitly.
This state of light is interesting because it may be created due to
interaction of one--mode light with the two--level system \cite{Ag.T.}.
The photon--coherent states of light $\vert  \alpha ,m\rangle$ have
the specific property since they do not contain the contribution of
states with fixed photon number $\vert n\rangle$ if the number $~n$
is less than $~m$. Thus the distribution function for photon--added
states $~p(n)$ is equal to zero up to the number $~n=m$.

The photon--added coherent states should be distinguished from the
states that are obtained by the action of the displacement operator on
the number state. These states have been studied in \cite{D1}, \cite{D2},
\cite{D3}. For these states the photon distribution function may be
nonzero for $~n<m$. One of mechanisms producing the squeezing and
correlation of the photon quadrature components is the parametric
excitation of the electromagnetic field oscillator \cite{T183}.
The parametric excitation may change the statistics of the photon--added
coherent states.

The aim of this work is to study the photon statistics of the photon
added states subjected to the parametric excitation. We obtain the
Wigner and $Q$--function of the parametrically excited photon added
states and study the deformations of photon distribution function of
squeezed and correlated light under the influence of the extra physical
parameter---the quantity of added photons. We will show that the
characteristic property of the distribution for photon--added coherent
states to be equal to zero up to a certain number of photons is
destroied by the parametric excitation of the electromagnetic field
oscillator.

\section{State $\vert \alpha,m \rangle$ of Stationary Oscillator }
%
Let us consider the state defined by the formula

\be           				\label{photon-added}
\vert \alpha,m \rangle =
N~exp\ds(-\frac{\vert \alpha \vert^2}{2})
\frac{\partial^m}{\partial \alpha^m}
\left(~exp(\frac{\vert\alpha\vert^2}{2})\vert\alpha\rangle
\right),
\ee
where $N=(m!L_m(-\vert\alpha\vert^2))^{-1/2}$
is a normalization constant, $\vert\alpha\rangle$ is a
coherent state \cite{Glauber} and $m$ is a positive integer.
The state $\vert\alpha\rangle$ can be defined as
\be					\label{coherent}
\vert\alpha\rangle =
exp(-\frac{\vert\alpha\vert^2}{2})
exp(\alpha\hat a^+)\vert 0 \rangle,
\ee
where $\vert 0 \rangle$ is a vacuum state and $\hat a,\hat a^+$
are photon creation and annihilation operators. Using this series
we obtain
\be					\label{photon-added1}
\vert\alpha,m\rangle =
\frac{\hat a^{+m}\vert\alpha\rangle}
{(m!~L_m(-\vert\alpha\vert^2))^{1/2}}.
\ee
This state coincides with the state $\vert\alpha,m\rangle$
that has been studied in \cite{Ag.T.}.
With the aid of (\ref{photon-added}) and using the formula for the
Hermite polynomial
\be
{}~H_m(\xi) = 				\label{Hermit1}
(-1)^m e^{\xi^2}\frac{d^m}{d\xi^m}
e^{-\xi^2},
\ee
we can easily write the evident expression for wave function of the
state $\vert\alpha,m\rangle$ in the coordinate representation
\be				\label{evident-ph-added}
\langle q\vert\alpha,m\rangle =
\frac{2^{-\frac{m}{2}}~H_m(q-\alpha/\sqrt 2)\langle q\mid\alpha\rangle}
{(m!~L_m(-\mid \alpha\mid^2))^{1/2}}.
\ee
The states $\vert\alpha,m\rangle$ should be distinguished from the
states $\vert n,\alpha\rangle$  that are defined as
\bdm
\vert n,\alpha\rangle=
D(\alpha)\vert n\rangle.
\edm
These states are related to the shifted harmonic oscillators. The
properties and statistics of these states were studied in \cite{D1,D2,D3}.
It easy to notice that the state $\vert n,\alpha\rangle$ is the
superposition of the states $\vert\alpha,m\rangle$
\bdm
\vert n,\alpha\rangle=
\frac{1}{\sqrt{n!}}\sum_{m=0}^nC_n^m
(-\alpha^*)^{n-m}(m!L_m(-\vert\alpha\vert^2))^{1/2}\vert\alpha,m\rangle.
\edm
The reciprocal expression can be derived as well
\bdm
\vert\alpha,m\rangle=(m!L_m(-\vert\alpha\vert^2))^{-1/2}
\sum_{n=0}^mC_m^n(\alpha^*)^{m-n}\sqrt{n!}\vert n,\alpha\rangle.
\edm
The photon distribution function for the state $\vert\alpha,m\rangle$
has been found in \cite{Ag.T.}
\be				\label{dsfnc}
p(n)=
\frac{n!\vert\alpha\vert^{2(n-m)}exp(-\vert\alpha\vert^2)}
{[(n-m)!]^2L_m(-\vert\alpha\vert^2)m!}.
\ee
We could calculate the generating function for this distribution in
the form
\be			\label{genfnc}
G(\lambda)=\frac{e^{(\lambda-1)\vert\alpha\vert^2}\lambda^m
L_m(-\lambda\vert\alpha\vert^2)}{L_m(-\vert\alpha\vert^2)},
\ee
which is associated with the expression (\ref{dsfnc}) by the formula
\be			\label{distGen}
G(z)=\sum_{n=0}^\infty p(n)z^n.
\ee
Below this expression will be generalized for the nonstationary case.
\section {Time--Dependent Photon--Added Coherent States}
We introduce the state $\vert\alpha,m,t\rangle$ which is ananlogous to
the state $\vert\alpha,m\rangle$ of stationary oscillator, but the new
state satisfies the Schr\"odinger equation with the Hamiltonian
\bdm
\hat H=\frac{\hat p^2}{2}+~\Omega^2(t)\frac{\hat q^2}{2}.
\edm
The initial value for the frequency $~\Omega(t)$ is taken to be
$~\Omega(0)=1$. We choose the initial state of the parametric oscillator
to be photon--added state $\vert\alpha,m,0\rangle=
\vert\alpha,m\rangle$, and we will study its evolution. In terms of
evolution operator $~\hat U(t)$ of the oscillator it means that
\be				\label{Ua}
\vert\alpha,m,t\rangle=~\hat U(t)\vert\alpha,m,0\rangle=
{}~\hat U(t)\vert\alpha,m\rangle,
{}~~~~~~\hat U(0)=~\hat 1.
\ee
The evolution operator $~\hat U(t)$ is unitary one,
\be				\label{UU+}
{}~\hat U(t)~\hat U^+(t)=~\hat 1.
\ee
Thus we can write
\be 					\label{UaU+}
\vert\alpha,m,t\rangle\sim
{}~\hat U(t)\hat a^{+m}~\hat U^+(t)~\hat U(t)\vert\alpha\rangle=
{}~\hat A^{+m}(t)\vert\alpha,t\rangle,
\ee
where $\vert\alpha,t\rangle$ is a time--dependent coherent state
\cite{Malkin.Man'ko}, and
\be					\label{A-t}
{}~\hat A^+(t)=~\hat U(t)~\hat a^+~\hat U^+(t)
\ee
is the integral of motion of the parametric oscillator.
{}From (\ref{Ua})  and (\ref{UU+}) we can derive that $~\hat A^+(t)$
satisfies the initial condition
\be					\label{init-cond}
{}~\hat A^+(0)=~\hat a^+.
\ee
By using hermitian conjugation of (\ref{A-t}) and (\ref{init-cond})
we get another integral of motion
\be					\label{ini-cond1}
{}~\hat A(t)=~\hat U(t)~\hat a~\hat U^+(t),
{}~~~~~~\hat A(0)=~\hat a.
\ee
Due to unitarity property of the evolution operator for hermitian
Hamiltonian, the operators $~\hat A(t)$ and $~\hat A^+(t)$
satisfy to the same commutation relation as $a$ and $a^+$
\be				\label{commutation-rel}
[~\hat A(t),~\hat A^+(t)]=~\hat 1.
\ee
Thus the algebra defined by the operators $~\hat A(t),~\hat A^+(t),
{}~\hat 1$ is the same Heisenberg--Weyl algebra as in the case
of operators $~\hat a,~\hat a^+,~\hat 1$. Using the definition of
invariants we get \cite{Malkin.Man'ko}

\be 					\label{Aqp-A+qp}
{}~\hat A(t)=\frac{i}{\sqrt 2}
(~\varepsilon(t)\hat p-\dot\varepsilon(t)q),
{}~~~~~~\hat A^+(t)=-\frac{i}{\sqrt 2}
(~\varepsilon^*(t)p-~\dot\varepsilon^*(t)q)
\ee
where $c$--number function $\varepsilon(t)$ satisfies the equation
\be			\label{e-equation}
{}~\ddot\varepsilon(t)+~\Omega^2(t)~\varepsilon(t)=0,
\ee
with the initial conditions $~\varepsilon(0)=1, ~~~\dot\varepsilon(0)=i,~~$
which means that the Wronskian is
\be				\label{Vronscian}
\varepsilon\dot\varepsilon^*-
\varepsilon^*\dot\varepsilon=-2i.
\ee
Thus, the solutions to Scr\"odinger equation for the arbitrary
time--dependence of the frequency are expressed in terms of the solution
$~\varepsilon(t)$ of the classical equation of motion (\ref{e-equation}).
Let us write down the evident expression for the state
$\vert\alpha,m,t\rangle$ in the coordinate representation
\be
\langle q\vert\alpha,m,t\rangle=
\frac{\left(\ds{\frac{\varepsilon^*}{2\varepsilon}}\right)^{\frac{m}{2}}
{}~H_m\left(\ds{\frac{q}{\vert\varepsilon\vert}-\sqrt{\frac{\varepsilon^*}
{2\varepsilon}}\alpha}\right)\langle q\vert\alpha,t\rangle}
{(m!~L_m(-\vert\alpha\vert^2))^{1/2}},
\ee
where $\langle q\vert\alpha,t\rangle$ is the time-dependent coherent state
\be				\label{timeCohSt}
\langle q\vert\alpha,t\rangle =
\pi^{-1/4}\varepsilon ^{-1/2}
exp(\frac{i\dot\varepsilon q^2}{2\varepsilon}+
\frac{\sqrt 2\alpha q}{\varepsilon}
-\frac{\alpha^2\varepsilon^*}{2\varepsilon}
-\frac{\vert\alpha\vert^2}{2}).
\ee
We see that the wave function of parametrically excited photon--added coherent
state is expressed in terms of Hermite polynomials.
\section {First and Second Quadrature Component Moments}
It is easy to derive from (\ref{Aqp-A+qp}) the expressions for photon
quadrature components in terms of the integral of motion
\be				\label{pq-of-AA+}
\hat p=\frac{1}{\sqrt 2}
(~A(t)\dot\varepsilon^*(t)+~A^+(t)\dot\varepsilon(t)),~~~~~~
\hat q=\frac{1}{\sqrt 2}
(~A(t)\varepsilon^*(t)+~A^+(t)\varepsilon(t)).
\ee
So, the problem of finding the moments of the quadratures is reduced
to finding the average values of products of the operators $~A$
and $~A^+$ in the states $~\vert\alpha,m,t\rangle$.
Using the results \cite{Ag.T.} we get
\be				\label{average-q}
\langle\hat q\rangle=
\langle\alpha,m,t\vert\hat q\vert\alpha,m,t\rangle=
\frac{~L_m^{(1)}(-\vert\alpha\vert^2)}
{\sqrt 2~L_m(-\vert\alpha\vert^2)}
(\varepsilon^*\alpha+\varepsilon\alpha^*),
\ee
\be				\label{average-p}
\langle\hat p\rangle=
\frac{~L_m^{(1)}(-\vert\alpha\vert^2)}
{\sqrt 2~L_m(-\vert\alpha\vert^2)}
(\dot\varepsilon^*\alpha+\dot\varepsilon\alpha^*),
\ee
\be				\label{average-q2}
\langle\hat q^2\rangle=
\frac{(\varepsilon^{*2}\alpha^2+\varepsilon^2\alpha^{*2})
{}~L_m^{(2)}(-\vert\alpha\vert^2)+
2\vert\varepsilon\vert^2(m+1)~L_{m+1}(-\vert\alpha\vert^2)-
\vert\varepsilon\vert^2~L_m(-\vert\alpha\vert^2)}
{2~L_m(-\vert\alpha\vert^2)},
\ee
\be				\label{average-p2}
\langle\hat p^2\rangle=
\frac{(\dot\varepsilon^{*2}\alpha^2+\dot\varepsilon^2\alpha^{*2})
{}~L_m^{(2)}(-\vert\alpha\vert^2)+
2\vert\dot\varepsilon\vert^2(m+1)~L_{m+1}(-\vert\alpha\vert^2)-
\vert\dot\varepsilon\vert^2~L_m(-\vert\alpha\vert^2)}
{2~L_m(-\vert\alpha\vert^2)}.
\ee
With the help of these expressions it is easy to write dispersions
of the quadratures $~\hat q$ and $~\hat p$
\be					\label{disp_q_p}
\sigma_q^2(t)=\langle\hat q^2\rangle-\langle\hat q\rangle^2,~~~~~~
\sigma_p^2(t)=\langle\hat p^2\rangle-\langle\hat p\rangle^2.
\ee
All the moments and dispersions are the functions of time as well
as of the amplitude $\vert\alpha\vert$ and the phase of $\alpha$. In
the stationary case the results coincide with the results obtained in
Ref. \cite{Ag.T.}.
\section {Mean and Dispersion of Photon Number}
Let us express the creation and annihilation operators of photons in
terms of the integrals of motion. We get
\be				\label{a-a+}
a=\frac{1}{2}(A(\varepsilon^*+i\dot\varepsilon^*)+
A^+(\varepsilon+i\dot\varepsilon)),~~~~
a^+=\frac{1}{2}(A^+(\varepsilon-i\dot\varepsilon)+
A(\varepsilon^*-i\dot\varepsilon^*)).
\ee
The average number of photons in the state $\vert \alpha ,~m,~t \rangle$
is
\bdm
\langle N\rangle=\langle\alpha,m,t\vert a^+a\vert\alpha,m,t\rangle=
(\vert\varepsilon\vert^2+
\vert\dot\varepsilon\vert^2)\frac{(m+1)~L_{m+1}(-\vert\alpha\vert^2)}
{2~L_m(-\vert\alpha\vert^2)}
\edm
\be				\label{average-a+a}
-\frac{1}{4}(\vert\varepsilon\vert^2+\vert\dot\varepsilon\vert^2)
-\frac{1}{2}+
\frac{L_m^{(2)}(-\vert\alpha\vert^2)}{4~L_m(-\vert\alpha\vert^2)}
(\alpha^{*2}(\varepsilon^2+\dot\varepsilon^2)+
\alpha^2(\varepsilon^{*2}+\dot\varepsilon^{*2})),
\ee
and the average of number of photons squared is given by the formula
\bdm
\langle (a^+a)^2\rangle=
(\alpha^{*4}(\varepsilon^2+\dot\varepsilon^2)^2+
\alpha^4(\varepsilon^{*2}+\dot\varepsilon^{*2})^2)
\frac{L_m^{(4)}(-\vert\alpha\vert^2)}
{16~L_m(-\vert\alpha\vert^2)}
\edm
\bdm
+(\vert\varepsilon\vert^2+\vert\dot\varepsilon\vert^2)(m+1)
(\alpha^{*2}(\varepsilon^2+\dot\varepsilon^2)+
\alpha^2(\varepsilon^{*2}+\dot\varepsilon^{*2}))
\frac{L_{m+1}^{(2)}(-\vert\alpha\vert^2)}
{4~L_m(-\vert\alpha\vert^2)}
\edm
\bdm
+(m+1)(m+2)(\vert\varepsilon^2+\dot\varepsilon^2\vert^2+
2(\vert\varepsilon\vert^2+\vert\dot\varepsilon\vert^2)^2)
\frac{L_{m+2}(-\vert\alpha\vert^2)}
{8~L_m(-\vert\alpha\vert^2)}
\edm
\bdm
-(m+1)(2(\vert\varepsilon\vert^2+\vert\dot\varepsilon\vert^2)
(\vert\varepsilon\vert^2+\vert\dot\varepsilon\vert^2+1)+
\vert\varepsilon^2+\dot\varepsilon^2\vert^2)
\frac{L_{m+1}(-\vert\alpha\vert^2)}
{4~L_m(-\vert\alpha\vert^2)}
\edm
\bdm
-(3(\vert\varepsilon\vert^2+\vert\dot\varepsilon\vert^2)+2)
(\alpha^{*2}(\varepsilon^2+\dot\varepsilon^2)+
\alpha^2(\varepsilon^{*2}+\dot\varepsilon^{*2}))
\frac{L_m^{(2)}(-\vert\alpha\vert^2)}
{8~L_m(-\vert\alpha\vert^2)}
\edm
\be					\label{average-a+aa+a}
+\frac{1}{16}((\vert\varepsilon\vert^2+\vert\dot\varepsilon\vert^2+2)^2+
2\vert\varepsilon^2+\dot\varepsilon^2\vert^2).
\ee
With the help of these expressions one can write down the expression
for the function which describes the dispersion of the number of photons
\cite{L.Mandel}
\be			\label{Q_of_a_and_m}
Q(\alpha,m)=\frac{\langle (a^+a)^2\rangle-
\langle a^+a\rangle^2}{\langle a^+a\rangle}.
\ee
It's worth noticing also that the average values of products of any number
of operators $A^n$ and $A^{+m}$ (where $m$ and $n$ are the arbitrary
integers) could be evaluated from the commutation relations.
\section {Time--Dependent Photon Distribution Function }
Next we study the photon distribution of the field in the state
$\vert\alpha,m,t\rangle$. By definition we could write for this
distribution fuction the expression
\bdm
p(n,t)=\vert\langle n\vert\alpha,m,t\rangle\vert^2.
\edm
To calculate the integral $\langle n\vert\alpha,m,t\rangle$ we will
use the evident expression for the states $\vert n\rangle$ and
$\vert\alpha,m,t\rangle$ in the coordinate representation, in which
the Hermite polynomials are taken in the following form
\be				\label{GenH}
H_m(\xi)=\lim_{\beta\to0}\frac{\partial^m}{\partial\beta^m}
exp(-\beta^2+2\beta\xi).
\ee
Since $\beta$ is an independent parameter to find the projection of the
state $\vert\alpha,m,t\rangle$ on the vector $\vert n\rangle$ means to
calculate the Gaussian integral. Using the Hermite polynomials of two
variables, that are defined by means of the generating function (see,
for example \cite{T183}),
\be			\label{GenHnm}
exp(-\frac{1}{2}\vec a\hat R\vec a+\vec a\hat R\vec y)=
\sum_{n,m=0}^\infty\frac{a_1^na_2^m}{n!m!}H_{nm}^{\lbrace R\rbrace}(y_1,y_2).
\ee
we get the distribution function in the form
\bdm
p(n,t)=
\frac{2\vert H_{mn}^{\lbrace R\rbrace}\left(-i\varepsilon^*\alpha/
\vert\varepsilon\vert,0\right)\vert^2}
{n!m!(\vert\varepsilon\vert^2+\vert\dot\varepsilon\vert^2+2)^{1/2}
{}~L_m(-\vert\alpha\vert^2)}
\edm
\be				\label{distribution-function}
\otimes exp\left(-\vert\alpha\vert^2
-Re\left(\frac{\varepsilon^{*2}+\dot\varepsilon^{*2}}
{\vert\varepsilon\vert^2
+\vert\dot\varepsilon\vert^2+2}\alpha^2\right)\right),
\ee
where $H_{mn}^{\lbrace R\rbrace}(\xi,\eta)$ is a Hermite polynomial
of two variables defined by the $2\times 2-$matrix $~R$ with the
following matrix elements
\bdm
R_{11}=\frac{\varepsilon}{\varepsilon^*}
\frac{i\dot\varepsilon^*-\varepsilon}{\varepsilon-i\dot\varepsilon},
{}~~~~~~R_{22}
=\frac{\varepsilon+i\dot\varepsilon}{\varepsilon-i\dot\varepsilon},
\edm
\bdm
R_{12}=R_{21}=\frac{2\varepsilon}{(\varepsilon-i\dot\varepsilon)
\vert\varepsilon\vert}.
\edm
In the stationary case the matrix $R$ reduces to
\bdm
R~~\to~~\left(_{1~~0}^{0~~1}\right),
\edm
and if taking into account that \cite{T183}
\bdm
H_{mn}^{\lbrace 00\rbrace}(\xi,\eta)=
(-1)^{\mu_{mn}}\mu_{mn}!\xi^{\ds(n-m+\vert n-m\vert)/2}
{}~~\eta^{\ds(m-n+\vert m-n\vert)/2}
{}~~L_{\mu_{mn}}^{\vert m-n\vert}(\xi\eta),
\edm
where
\bdm
\mu_{mn}=min(m,n),
\edm
we reproduce the result of Ref. \cite{Ag.T.} for stationary case.

Now we write down the generating function, which is defined as
\be			\label{defGenF}
G(z)=\sum_{n=0}^\infty p(n,t)z^n.
\ee
Calculating the expression for the state
$\vert\alpha,m,t\rangle$ (see above) we derived also a relationship
between the states $\vert\alpha,t\rangle$ and
$\vert\alpha,m,t\rangle.$
Therefore, it is easy to obtain the relation between
the projections of these states on the state
$\vert n\rangle$
\bdm
\langle n\vert\alpha,m,t\rangle =
(m!~L_m(-\vert\alpha\vert^2))^{-1/2}
exp(-\frac{\vert\alpha\vert^2}{2})
\frac{\partial^m}{\partial\alpha^m}
exp(\frac{\vert\alpha\vert^2}{2})
\langle n\vert\alpha,t\rangle.
\edm
Let us introduce the distribution function for $m=0.$
This will be
\be			\label{dsfm0}
p_0(n,t)=\vert\langle n\vert\alpha,t\rangle\vert^2.
\ee
Then the relation between the functions $p(n,t)$ and $p_0(n,t)$
is given by the formula
\be		\label{pp0}
p(n,t)=(m!~L_m(-\vert\alpha\vert^2))^{-1}
e^{-\vert\alpha\vert^2}\frac{\partial^{2m}}
{\partial\alpha^m\partial\alpha^{*m}}
e^{\vert\alpha\vert^2}p_0(n,t).
\ee
The same relation exists for the generating functions as well. Due to this
to find $G(z,t)$ it is enough to compute the function
\be			\label{G0}
G^0(z,t)=\sum_{n=0}^\infty p_0(n,t)z^n.
\ee
The function $p_0$ is proportional to the $\vert H_n\vert^2,$
therefore, we need to obtain the relation for summing the modules square
of the Hermite polynomials. To obtain it we consider the product of two
generating functions for the
Hermite polinomials
\be			\label{HnHm}
exp(-x^2+2x\xi)exp(-y^2+2y\xi^*)=
\sum_{n=0}^\infty\sum_{m=0}^\infty\frac{x^n}{n!}\frac{y^m}{m!}
H_n(\xi)H_m(\xi^*),
\ee
making a substitution of variables
\bdm
x=\sqrt\lambda\chi,~~~~~~y=\sqrt\lambda\chi^*,
\edm
where $\lambda$ is a real parameter.
Then multiplying both sides of (\ref{HnHm}) by the factor
$exp(-\chi\chi^*)$ and calculating the Gaussian integral with the
account of the relation
\bdm
\frac{i}{2\pi}\int\chi^{*n}\chi^m exp(-\chi\chi^*)d\chi d\chi^*=
n!\delta_{nm},
\edm
we obtain the expression for $G^0(z,t),$ where the relation
between $\lambda$ and $z$ is given by the following formula
\bdm
\lambda=\frac{z\vert\varepsilon+i\dot\varepsilon\vert}
{2\vert\varepsilon-i\dot\varepsilon\vert}~~~.
\edm
Using (\ref{pp0}) we find the expression for the generating
function
\be				\label{genFuncOft}
G(z,t)=\frac{\sqrt\sigma_0}{m!L_m(-\vert\alpha\vert^2)}
H_{mm}^{\lbrace J\rbrace}(\alpha,\alpha^*)
exp[(z\sigma_0-1)\vert\alpha\vert^2-Re(J_{11}\alpha^2)].
\ee
where $J$ is $2\times 2$-matrix with the following matrix elements
\bdm
J_{11}=J_{22}^*=\frac{2\varepsilon^*-[\varepsilon^*+i\dot\varepsilon^*+
(\varepsilon^*-i\dot\varepsilon^*)z^2]\sigma_0}{2\varepsilon},
\edm
\bdm
J_{12}=J_{21}=-z\sigma_0,
\edm
\bdm
\sigma_0=\frac{4}{(\vert\varepsilon\vert^2+\vert\dot\varepsilon\vert^2)
(1-z^2)+2(1+z^2)}~~~.
\edm
One can check that $G(1,t)=1$. From the definition (\ref{defGenF}) it
follows that
\bdm
\langle a^+a\rangle=\lim_{z\to1}\frac{\partial G(z,t)}{\partial z}~~~.
\edm
Taking the limit $z\to1$ we see that
this result coincides with the expression (\ref{average-a+a}) for
$\langle a^+a\rangle$ obtained above.
\section {Quasiprobability Distributions for Field
in $\vert\alpha,m,t\rangle$--State}
Next we study the $Q$-function which is the diagonal matrix elements of
the density operator in coherent states basis (see, for example,
\cite{Mehta.Sudarshan}).
Thus, the $Q$-function is defined by formula
\be			\label{Q-function}
Q(z)=\langle z\vert\alpha,m,t\rangle
\langle\alpha,m,t\vert z\rangle.
\ee
Let us calculate the matrix element $\langle z\vert\alpha,m,t\rangle$.
As in the case of finding the distribution function the Hermite
polynomials can be expressed through the generating function. The
calculation gives
\bdm
Q(z)=\frac{exp(-\vert z\vert^2-\vert\alpha\vert^2)
2\vert\vartheta\vert^{2m}\vert H_m(\varsigma)\vert^2}
{L_m(-\vert\alpha\vert^2)m!(\vert\varepsilon\vert^2+
\vert\dot \varepsilon\vert^2+2)^{1/2}}
\edm
\be			\label{evidant_Q-function}
\times exp\left(Re~\frac{(\varepsilon+i\dot\varepsilon)z^{*2}+
4z^*\alpha-(\varepsilon^*-i\dot\varepsilon^*)\alpha^2}
{\varepsilon-i\dot\varepsilon}
\right),
\ee
where
\bdm
\vartheta=\left(\frac{\varepsilon^*-i\dot\varepsilon^*}
{2(\varepsilon-i\dot\varepsilon)}\right)^{1/2},
{}~~\varsigma=\vartheta\left(\frac{2z^*}{\varepsilon^*-i\dot\varepsilon^*}
-\alpha\right).
\edm
A transition to the stationary case is obvious. Preexponential factor
in the formula (\ref{evidant_Q-function}) is related to $~m$ added photons
and it changes the Gaussian form of the $~Q$--function of the coherent
state described by the exponent. Thus, tjhe $~Q$--function of the
photon--added coherent state has the shape of deformed Gaussian with
extra maxima and minima.

The Wigner function of an arbitrarily state, determined by
the density matrix $\rho(q,q')$, is defined by the relation
\be					\label{defW}
W(q,p)=\int\limits_{-\infty}^{+\infty}
\rho(q+\frac{r}{2},q-\frac{r}{2})exp(-ipr)dr.
\ee
For the pure state $\vert\alpha,m,t\rangle$ we have
\bdm
\rho(q,q')=\langle q\vert\alpha,m,t\rangle\langle\alpha,m,t\vert q'\rangle.
\edm
Therefore, to calculate the integral
(\ref{defW}) we make the same computation as for the integral
$\langle n\vert\alpha,m,t\rangle$ found above. For two-dimensional
Hermite polynomial having zero diagonal matrix elements of the matrix $R$
the following relation holds

\bdm
H_{nn}^{\lbrace00\rbrace}(\xi,\eta)=
(-1)^n n!L_n(\xi\eta),~~~~~~~~~R=\left(_{1~~0}^{0~~1}\right),
\edm
Due to this we can represent the Wigner function in the form
\be			\label{Wigner}
W(q,p)=\frac{(-1)^m2L_m(\vert\gamma\vert^2)}{L_m(-\vert\alpha\vert^2)}
exp(-\vert i\varepsilon p-i\dot\varepsilon x-\sqrt 2\alpha\vert^2),
\ee
where $\gamma$ is defined by
\bdm
\gamma=ip\vert\varepsilon\vert\sqrt 2-i\sqrt 2
\frac{\dot\varepsilon\varepsilon^*}
{\vert\varepsilon\vert}q-\frac{\alpha\varepsilon^*}
{\vert\varepsilon\vert}.
\edm
The physical parameters $~m$ describing extra photons produces the
factor $~L_m(\vert \gamma \vert ^{2})$ in the formula (\ref{Wigner})
and this factor changes the Gaussian shape of the Wigner function of
the coherent state. Since Laguerre polynomial has the wavy behaviour
the Wigner function of photon--added state has local maxima and minima.
Thus, we obtained the phase space distributions for the photon--added
states of the parametric oscillator.

\section {Numerical Evaluation of Means and Dispersions for
Specific Time-Dependence of Frequency}
In this Section we consider a spesific time-dependence of the oscillator
frequency $\Omega^2(t)$ in form of "the saw with one tooth",i.e.,
\bdm
\Omega^2(t)=1,~~if~~t\o 0~~or~~t\geq T,
\edm
and
\bdm
\Omega^2(t)=1+kt,~~if~~0<t<T.
\edm
The time-dependence is determined by two parameters $~T$ and $~k$.
Function $\varepsilon(t)$ for such a "saw" is given by the formula
\bdm
\varepsilon(t)=e^{it},~~~~if~~t<0,
\edm
\bdm
\varepsilon(t)=
\sqrt{1+kt}(C_1J_{1/3}(\frac{2(1+kt)^{3/2}}{3k})+
C_2Y_{1/3}(\frac{2(1+kt)^{3/2}}{3k})),~~~if~~0<t<T,
\edm
where $C_1$ and $C_2$ are defined by
\bdm
C_1=\frac{Y_{4/3}(\frac{2}{3k})-Y_{1/3}(\frac{2}{3k})(k-i)}
{J_{1/3}(\frac{2}{3k})Y_{4/3}(\frac{2}{3k})-
J_{4/3}(\frac{2}{3k})Y_{1/3}(\frac{2}{3k})},
\edm
\bdm
C_2=\frac{J_{1/3}(\frac{2}{3k})(k-i)-J_{4/3}(\frac{2}{3k})}
{J_{1/3}(\frac{2}{3k})Y_{4/3}(\frac{2}{3k})-
J_{4/3}(\frac{2}{3k})Y_{1/3}(\frac{2}{3k})},
\edm
and if $t\geq T$, then $\varepsilon (t)$-function is given by
\bdm
\varepsilon=D_1e^{it}+D_2e^{-it},
\edm
where $D_1$ and $D_2$ are given by
\bdm
D_1=\frac{e^{-iT}}{2}(\varepsilon(T)-i\dot\varepsilon(T)),
\edm
\bdm
D_2=\frac{e^{iT}}{2}(\varepsilon(T)+i\dot\varepsilon(T)).
\edm
At the fixed perturbation parameters $k$ and $T$ the dispersions
of quadratures $\hat q$ and $\hat p$ will be periodical functions of
time when $t>T$. The minimal value of dispersion squared multiplied by
the factor 2 is the squeezing coefficient $s$
\bdm
s=\frac{1}{2\sigma_q^2}.
\edm
We studied numerically the dependence of squeezing on
$k,\alpha,m$ at the fixed $T=1$. In Fig. 1 there are presented
the results of the numerical calculation for parameter $\alpha =10$,
which corresponds to the multiphoton field. Diferent curves correspond
to diferent $m$. We can see that they are very close. Their difference
can be seen better at $k\sim 30$. In this range the values of
$m=1,5,10,50$ correspond to the curves, if one counts from top to bottom.
The dependence of squeezing on $k$ is an oscillating function. When
$k\approx 60$ we observe the local minimum of squeezing, which is equal
approximately to $0.02$. In the following local minimum the squeezing
is even stronger.

The parameter $\alpha =1$ corresponds to the Fig. 2. In this case the
field state is intermediate between classical and quantum one. We can
see that the dependence on $m$ is stronger. In the range of $k$ from
$30$ to $80$, the upper curve correspons to $m=5$, and the lower one---
to $m=1$.

In Fig. 3 there are presented the plots for quantum state of the field
$\alpha =0.1$. The dependence on $m$ becomes even stronger. The curves
do not overlap as it  was in Figs. 1 and 2. The number $m=1$ corresponds
to the upper curve and $m=5$ to the lower one.

So we can conclude that with the decrease of $\vert\alpha\vert$
the field becomes more sensitive to photon pumping.

Now we discuss the behaviour of average number of photons and the
dispersions of a number of photons. When $t>T$ the Hamiltonian becomes
stationary and therefore the operator of a number of photons $a^+a$ is the
integral of motion. The average values of $\langle\hat N\rangle,~
\langle\hat N^2\rangle$ and therefore the dispersion of a number of
photons do not depend on time when $t>T$.
The dependences of the ratio of the dispersion of a number of photons
squared and mean photon number $\langle\hat N\rangle$ (which is denoted
$Q(\alpha ,m)$) on $k$ at the fixed $T=1,$ $t>T,$ $\alpha=1,$
$m=5$ is presented in Fig. 4. As in the case of squeezing
they have the tendency of the increasing role of $m$ when
$\vert\alpha\vert$ decreases. If we compare Fig. 4 with Fig. 2 we can
notice that when squeezing increases an average number of photons and the
dispersion of it in the field also increases.

Let us discuss the results of the distribution function investigation.
For the stationary case $t>T$ the distribution $p(n)$ does not depend
on time and the most interesting is its dependence on $n$ and $k$. In Fig. 5
$p(n)$ is ploted for $m=2$, $\alpha=1$. We chose the range of $k$ from
$50$ to $80$. According to Fig. 2 it is the range in which strong
squeezing is achieved. We can see here the known oscillations
of $p(n)$ \cite{Sch.W}, which proves nonclassical nature of the field.
At smaller sqeezing these oscillations are smaller, too. The dependence
of the photon distribution function on $k$ behaves as smoothly oscillating
(Fig. 5). While varying $k$, the behaviour of maxima and minima of $p(n)$
is changing. This is seen in Fig. 5 when $n$ varies from $15$ to $20$.

In Fig. 6 the dependence of the distribution function on $\alpha$ is
illustrated. Here we chose $k=63$, $T=1$, $m=2$, $t>T$ and $\alpha$ is real.
At $\alpha=0$ we get the distribution function of a squeezed Fock state
$\vert m,t\rangle$ \cite{Malkin.Man'ko} which looks in the coordinate
representation as
\bdm
\langle q\vert m,t\rangle =
\left(\frac{\dot\varepsilon^*}{2\varepsilon}\right)^{m/2}
(m!\varepsilon\sqrt\pi)^{-1/2}
exp\left(\frac{i\dot\varepsilon}{2\varepsilon}q^2\right)
H_m\left(\frac{q}{\vert\varepsilon\vert}\right).
\edm
While $\alpha$ is changing in the quantum range from $0$ to $1$,
the distribution oscillations can be easily noticed. The increase of
$\alpha$ results in the classical limit of a large number of photons.
It produced a strong delocalization of the distribution function of the
parameter $n$. Thus, the oscillations become smaller and invisible in
the graph.

The distribution function at $k=63$, $T=1$, $\alpha=1$, $t>T$ for $m=1$
and $m=2$ is shown in Fig. 7. For $m=1$ at $n>40$ oscillations disappear
while for $m=2$ they exist at large values of $n$.

Generaly speaking, the problem of dependence of the distribution function for
the squeezed and, simalteneously, correlated state on the parameter $m$
requires special investigation. We can just suppose that the trend of the
dependence on $m$ will be different at different extent of squeezing and
different correlation between quadrature components.

In Fig. 8 there is $Q$--function at $T=1$, $k=10$, $\alpha=1$, $m=5$, $t=10$.
If $\alpha$ is increased, the form of a $Q$--function peak will resemble
the form shown at fig.8. However, a shape and a position will be different.
Approximately, the same trend is observed while increasing $m$. At the values
of $\alpha$ corresponding to the quantum case the form of peak becomes much
more sensitive to changes of parameters. It is illustrated in
Fig. 9, which represents the $Q$--function at $\alpha=0.1$, $T=1$, $k=1$,
$m=10$, $t=10$.

The Wigner function is more sensitive to changes of $m$. In Fig. 10a there
is Wigner function at $T=1$, $k=10$, $\alpha=2$, $m=9$, $t=10$. The Wigner
function of the coherent state is the Gaussian function with one maximum.
On the other hand, fot the photon--added states such shape is deformed and
several local maxima are observable. Fig.10b shows the same plot that
Fig.10a but from the direction of $p$ axis. In this figure it is clearly
seen that the function $W(q,p)$ can be negative.

In Figs. 11 and 12 the Wigner function is ploted at the same values of
parameters $T=t=1$, $\alpha=2$, $m=1$ but at different values of $k$, $10$,
and $60$, respectively. At $k=10$ the dispersion of $q$ is greater than at
$k=60$ but the dispersion of $p$ is less. That is why the peak of the Wigner
function in Fig. 11 is significantly narrower along $q$ and wider along
$p$, than in Fig. 12.

Thus we demonstrated that parametrical excitation of the electromagnetic
field oscillator changes the statistics of photon--added coherent states
since the squeezing and correlation of quadratures produced due to
nonstationarity of the oscillator influences the dispersions and means
of photon numbers in these states.

\newpage
%
%
%

%
\newpage
\mbox{}~~~~~~~~~~~~~~~~~~~~~~~~~~~~~~~~~~~~~~~{\bf FIGURE CAPTIONS}\\

\hspace{1.5em}	\\
{\bf Fig. 1.} Squeezing of $q$ as a function
of $k$ for $T=1,$ $\alpha=10,$,
and $m=1,~5,~10,~50$.	\\

{\bf Fig. 2.} Squeezing of $q$ as a function
of $k$ for $T=1,$ $\alpha=1,$, and
$m=1,~5$.	\\

{\bf Fig. 3.} Squeezing of $q$ as a function
of $k$ for $T=1,$ $\alpha=0.1,$ and $m=1,~5$.	\\

{\bf Fig. 4.} The dispersion of the number of photons squared and
divided by the average number
of photons $Q(\alpha,m)$ for $T=1,$ $t>T,$ $\alpha=1,$ and $m=1,~5,~10.$ \\

{\bf Fig. 5.} Photon number distribution $p(n,t)$ as a function of
$n$ and $k$ for $T=1,$ $t>T,$ $\alpha=1,$ and $m=2.$ 	\\

{\bf Fig. 6.} Photon number distribution $p(n,t)$ as a function of
$n$ and $\alpha$ for $T=1,$ $t>T,$ $k=63,$ and $m=2$.	\\

{\bf Fig. 7.} Photon number distribution $p(n,t)$ for $T=1,$ $t>T,$
$k=63,$ $\alpha=1,$ and $m=1,~2$. 	\\

{\bf Fig. 8.} $Q$--function $Q(z)$ for $T=1,$ $t=10,$ $k=10,$ $\alpha=1,$
and $m=5.$ 	\\

{\bf Fig. 9.} $Q$--function $Q(z)$ for $T=1,$ $t=10,$ $k=10,$ $\alpha=0.1,$
and $m=10$. 	\\

{\bf Fig. 10a (b).} Wigner function $W(q,p)$ for $T=1,$ $t=10,$
$k=10,$ $\alpha=2,$ and $m=9$.	\\

{\bf Fig. 11.} Wigner function $W(q,p)$ for $T=1,$ $t=1,$
$k=10,$ $\alpha=2,$ and $m=1$.	\\

{\bf Fig. 12.} Wigner function $W(q,p)$ for $T=1,$ $t=1,$
$k=60,$ $\alpha=2,$ and $m=1$.	\\
\end{document}